\documentclass[a4paper,11pt]{article}
\pdfoutput=1 

\usepackage{jcappub} 

\usepackage[T1]{fontenc} 
\usepackage{bm}
\usepackage{graphicx}
\usepackage{hyperref}
\usepackage{xcolor}
\usepackage{multirow}
\usepackage{hhline}

\usepackage{array}

\definecolor{ShamrockGreen}{rgb}{0.0, 0.62, 0.38}

\newcommand{\update}[1]{{\leavevmode\color{ShamrockGreen}#1}}

\begin{document}

\title{Advancing Globular Cluster Constraints on the Axion-Photon Coupling}

\author{Matthew J. Dolan,}
\author[1]{Frederick J. Hiskens,\note{Corresponding author}}
\author{and Raymond R. Volkas}
\affiliation{ARC Centre of Excellence for Dark Matter Particle Physics, School of Physics, The University of Melbourne, Victoria 3010, Australia}

\emailAdd{dolan@unimelb.edu.au}
\emailAdd{fhiskens@student.unimelb.edu.au}
\emailAdd{raymondv@unimelb.edu.au}

\arxivnumber{2207.03102}

\date{\today}

\abstract{We improve the current upper bound on the axion-photon coupling derived from stellar evolution using  the $R_2$ parameter, the ratio of stellar populations on the Asymptotic Giant Branch to Horizontal Branch in Globular Clusters. We compare this with data from simulations using the stellar evolution code \texttt{MESA} which include the effects of axion production. Particular attention is given to quantifying in detail the effects of uncertainties on the $R$ and $R_2$ parameters due to the modelling of convective core boundaries. Using a semiconvective mixing scheme we constrain the axion-photon coupling to be $g_{a\gamma\gamma} <  0.47 \times 10^{-10}~\mathrm{GeV}^{-1}$. This rules out new regions of QCD axion  and axion-like particle parameter space. Complementary evidence from asteroseismology suggests that this could improve to as much as $g_{a\gamma\gamma} <  0.34 \times 10^{-10}~\mathrm{GeV}^{-1}$ as the uncertainties surrounding mixing across convective boundaries are better understood.}

\maketitle
\flushbottom

\section{Introduction}
Axions were first identified by Weinberg \& Wilczek as a prediction of the Peccei-Quinn mechanism which explains the absence of charge-parity (CP) violation in quantum chromodynamics (QCD), the so-called \textit{strong CP problem} \cite{PQ-PRD, PQ-PRL, Weinberg-40.223, Wilczek:1977pj}. The Peccei-Quinn mechanism introduces a new global, anomalous $U(1)_{\mathrm{PQ}}$ symmetry, which is spontaneously broken at energy scale $f_a$. The resulting pseudo Nambu-Goldstone boson - the axion - dynamically cancels the QCD-$\theta$ term, solving the problem. The motivation for axions is enhanced further by the fact that they can constitute dark matter~\cite{Preskill:1982cy, Abbott:1982af, Dine:1982ah}.

Though difficult to detect in terrestrial environments, axions could be produced copiously in the extreme conditions deep within stars \cite{Raffelt:1996wa}. Once produced, such axions would freely escape the star and drain energy from its interior. Stellar regions experiencing heightened energy-loss typically respond by contracting and heating. This intensifies nuclear burning therein, which expedites the associated evolutionary phase and can lead to a discrepancy between theory and observation. Constraints derived in this manner are termed \textit{stellar cooling} bounds. 

 Axions interact with photons via the two-photon interaction
\begin{equation}
    \label{eq: axion-photon L}
    \mathcal{L}_{a\gamma\gamma}=-\frac{g_{a\gamma\gamma}}{4}aF_{\mu\nu}\Tilde{F}^{\mu\nu},
\end{equation}
where $a$ is the axion field, and $F_{\mu\nu}$ ($\Tilde{F}^{\mu\nu}$) is the (dual) electromagnetic field strength tensor, and $g_{a\gamma\gamma}$ is the axion-photon coupling.  \textit{In this paper we provide new constraints on the axion-photon coupling $g_{a \gamma\gamma}$ using stellar evolution.}

\begin{figure*}[t]
    \centering
    \includegraphics[width = 0.7\textwidth]{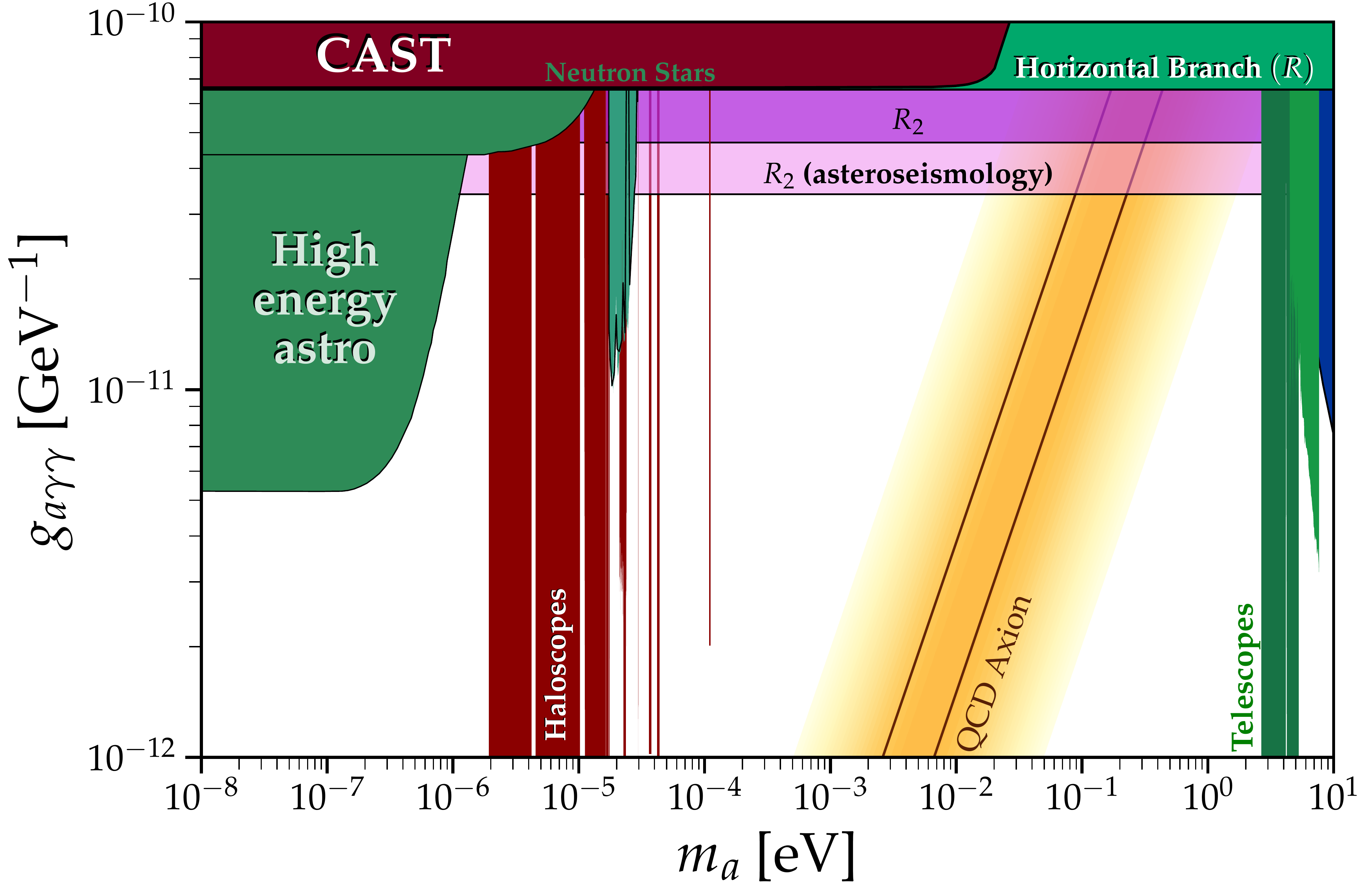}
\caption{The ALP parameter space and current experimental constraints, adapted from \cite{AxionLimits}, for $m_a$ between $10^{-8}$~eV and $10$~eV. The $R$-parameter constraint of Ref \cite{Ayala:2014pea} is labelled 'Horizontal Branch ($R$)', and the QCD axion band is shown as the yellow/orange strip. Other references are included in the text. Our most conservative new constraint is shown in purple, while the constraint supported by complementary evidence from asteroseismology is in pink.}
    \label{fig: param space}
\end{figure*}

The bound we derive also applies to  more general \textit{axion-like particles} (ALPs), which share the interaction of Equation \ref{eq: axion-photon L}. Unlike axions, ALPs do not solve the strong CP problem. While QCD axions have a linear relationship between the axion mass $m_a$ and the coupling $g_{a\gamma\gamma}$, ALPs  have independent values of $m_a$ and $g_{a\gamma\gamma}$. 
We show the ALP parameter space and current experimental constraints in Fig.~\ref{fig: param space} for $10^{-8} \leq m_a \leq 10$~eV. We have adapted the limits from~\cite{AxionLimits}.  This includes constraints from the CERN Axion Solar Telescope (CAST)~\cite{CAST:2017uph}, dark matter haloscopes~\cite{Ouellet:2018beu, Salemi:2021gck, Asztalos2010, ADMX:2018gho, ADMX:2019uok, ADMX:2021abc, ADMX:2018ogs, Bartram:2021ysp, Crisosto:2019fcj, Lee:2020cfj, Jeong:2020cwz, CAPP:2020utb, Grenet:2021vbb, HAYSTAC:2018rwy, HAYSTAC:2020kwv, McAllister:2017lkb, Alesini:2019ajt, Alesini:2020vny, CAST:2021add, DePanfilis, Gramolin:2020ict, Hagmann, Arza:2021rrm, Devlin:2021fpq, Thomson:2019aht}, cosmology~\cite{Cadamuro:2011fd} and a variety of astrophysical analyses, the most restrictive of which come from magnetic white dwarf polarisations~\cite{Dessert2022} and X-rays~\cite{Dessert:2021bkv}, neutron stars~\cite{Foster:2020pgt, Darling:2020uyo, Battye:2021yue}, the $R$-parameter of globular clusters~\cite{Ayala:2014pea}, the MUSE~\cite{Regis:2020fhw} and VIMOS~\cite{Grin:2006aw} spectroscopes and the Hubble Space Telescope~\cite{Nakayama:2022jza}. The region of the ALP plane inhabited by the (QCD) axion is indicated by the yellow/orange line, the width of which originates from different ultra-violet completions. The two red-lines within the QCD axion band define the canonical DFSZ~\cite{Dine:1981rt,Zhitnitsky:1980tq} and KSVZ~\cite{Kim:1979if,Shifman:1979if} models.

The previous stellar cooling bound labelled `Horizontal Branch ($R$)' in Fig. \ref{fig: param space} comes from the observed limits on the ratio of horizontal branch (HB) to red giant branch (RGB) stars in globular clusters, the so-called \textit{R-parameter} \cite{Ayala:2014pea}. Axions with $g_{10}=g_{a\gamma\gamma}/(10^{-10}\ \mathrm{GeV}^{-1})\sim 1$ are seldom produced in the degenerate cores and hydrogen burning shells of RGB stars, but can drain energy from the helium-burning cores of HB stars. Their inclusion in stellar models therefore reduces the duration of this evolutionary phase $\tau_{\mathrm{HB}}$ and lowers theoretical estimates for $R$, which if too efficient, can be excluded by observation.

While constraints on Beyond the Standard Model (BSM) particles have typically focused on the $R$-parameter, complementary bounds can be generated using other globular cluster parameters. The ratio of asymptotic giant branch (AGB) to horizontal branch stars (the $R_2$-parameter) is a particularly appealing candidate for this, as the helium-burning shell of AGB stars is hotter and less dense than HB cores. This makes it a potentially more sensitive probe of axion energy-loss, the rate of which scales as $T^7/\rho$~\cite{Dominguez1999, Dolan2021}. 
The authors of Ref \cite{Lattanzio2} recently established a more robust and self-consistent constraint of $R_2=0.117\pm0.005$ based on the Hubble Space Telescope photometry of 48 globular clusters.

In this paper we will develop limits on axion energy-loss from $R_2$ based on the results of Ref~\cite{Lattanzio2} and use these to extend existing constraints from $R$. This has the additional benefit of mitigating the impact of the treatment of mixing beyond convective boundaries, which is a major theoretical uncertainty on HB modelling, but has not been explored quantitatively in previous BSM constraints. $R_2$ is also known to be robust against variation in chemical composition, particularly the initial helium mass fraction $Y_{\mathrm{init}}$ \cite{Lattanzio2}. In contrast, it is well-known that a strong degeneracy exists between $Y_{\mathrm{init}}$ and $g_{10}$ when calculating $R$, which leads to significantly weakened constraints when large values of $Y_{\mathrm{init}}$ are adopted \cite{Ayala:2014pea}. We find that $g_{a\gamma\gamma}<0.47\times 10^{-10}\mathrm{GeV}^{-1}$. This rules out new regions of ALP parameter space for ALP masses between $\mathcal{O}(10^{-6})$ and $\mathcal{O}(1)$~eV. It also rules out a previously unconstrained part of the QCD axion parameter space for $m_a\sim 0.1$~eV, which was previously constrained by the $R$-parameter \cite{Ayala:2014pea}\footnote{Note that we are not considering constraints on the axion mass derived using the axion-electron coupling, e.g. those from the red giant branch tip luminosity \cite{RaffeltRGT, StranieroRGT}, as this interaction is not present in KSVZ-type models.} (see the intersection between the yellow/orange region and horizontal branch constraint in Fig.~\ref{fig: param space}). This new constraint is shown in purple in Fig.~\ref{fig: param space}, labelled `$R_2$'. There is also evidence from asteroseismology that would justify using the predictive mixing convective boundary scheme. This would lead to a bound of $g_{a\gamma\gamma}<0.34\times 10^{-10}\mathrm{GeV}^{-1}$. This constraint is shown in pink in Fig.~\ref{fig: param space}, labelled `$R_2$ (asteroseismology)', and we discuss this point further in Section~\ref{sec: GC constraint}.

 In Section \ref{sec: sec2}, we describe our simulations, including the impact of uncertainties from mixing across convective boundaries. We then describe the manner in which $R$ and $R_2$ can be calculated from the output of stellar models in Section \ref{sec: GC constraint} and present our overall constraint, before before summarising and concluding in Section \ref{sec: conclusion}. Additional details pertaining to the treatment of axion energy-loss in our stellar evolution models, our choice of input physics, the importance of different schemes for modelling convective boundaries and additional details about our calculations for $R$ and $R_2$ can be found in Appendices \ref{app: axion energy-loss}, \ref{app: simulations}, \ref{sec: CB modelling} and \ref{sec: Calculation details} respectively.

\section{Simulating globular cluster stars}
\label{sec: sec2}

The construction of robust bounds on axions from either $R$ or $R_2$ hinges upon our ability to reliably simulate globular cluster stars. Here we describe our simulations, which employ \textit{Modules for Experiments in Stellar Astrophysics} (\texttt{MESA}), a state-of-the-art open source, 1D stellar evolution code \cite{MESA1, MESA2, MESA3, MESA4, MESA5}.

\subsection{Simulations}
\label{sec: simulations}
We use an edited version of \texttt{MESA}, which we have extended to include energy-loss to axions which are sufficiently light to be thermally produced in HB cores (less than a few keV). We do so following the method outlined in Ref \cite{Raffelt:1987yu}, which includes the effects of electron degeneracy necessary to describe energy-loss in RGB stars. Details of this implementation are provided in Appendix \ref{app: axion energy-loss}. The primary production mechanism of such axions is the \textit{Primakoff process} ($Ze+\gamma\rightarrow Ze+a$), the conversion of a thermal photon into an axion in the presence of an external electromagnetic field provided by the constituent charges in the stellar plasma. 

Our simulations are separated into three distinct phases. In the first of these, we evolve a star of initial mass $M_{\mathrm{init}}$, metallicity $Z$ and initial helium mass fraction $Y_{\mathrm{init}}$ from the pre-Main Sequence until the zero-age horizontal branch (ZAHB), when stable helium burning begins in the core. Unless otherwise specified we adopt values of $M_{\mathrm{init}}=0.82M_{\odot}$ and $Z=0.001$, which are appropriate choices for HB stars in globular clusters\footnote{The dependence of the limits from $R$ on $M_{\mathrm{init}}$ and $Z$ has been studied in Ref~\cite{Ayala:2014pea} who found it had a minor impact, though quantitative information is not available. The influence of these parameters on $R_2$ was similarly studied in Ref~\cite{Lattanzio2}, where it was found that $\frac{\partial R_2}{\partial\log_{10} Z}\approx-0.003$ and $\frac{\partial R_2}{\partial M_{\mathrm{init}}}<-0.07$. The resulting variation in $R_2$ is thus less than a few percent for the ranges of these parameters appropriate for globular clusters ($\Delta\log_{10} Z < 1$ and $-0.02<\Delta M_{\mathrm{init}}/M_{\odot}<0.03$). Consequently we do not vary $M_{\mathrm{init}}$ or $Z$.}.

Our second set of simulations run from the ZAHB to the terminal-age horizontal branch (TAHB), when the central helium mass fraction $Y_c$ falls below $10^{-4}$. For globular cluster stars, the HB is the first evolutionary phase for which the core is convective. Consequently, these simulations must include an appropriate choice for modelling mixing beyond convective boundaries (CBs), the significance of which is discussed in Section \ref{sec: CBs}. These simulations can suffer from a lack of convergence when different temporal and spatial resolutions are employed \cite{MESA4}. Consequently, they are performed a number of times with different maximum timesteps and model sizes to sample any stochastic variation present.

Finally, once the TAHB has been reached, we switch off the chosen convective boundary scheme and allow the simulations to run through the asymptotic giant branch phase, which immediately follows the HB, until mass loss has reduced the outer envelope to only 1\% of the total stellar mass. Detail of our adopted input physics can be found in Appendix \ref{app: simulations}.

\begin{figure*}[t!]
    \centering
    \includegraphics[width=0.47\textwidth]{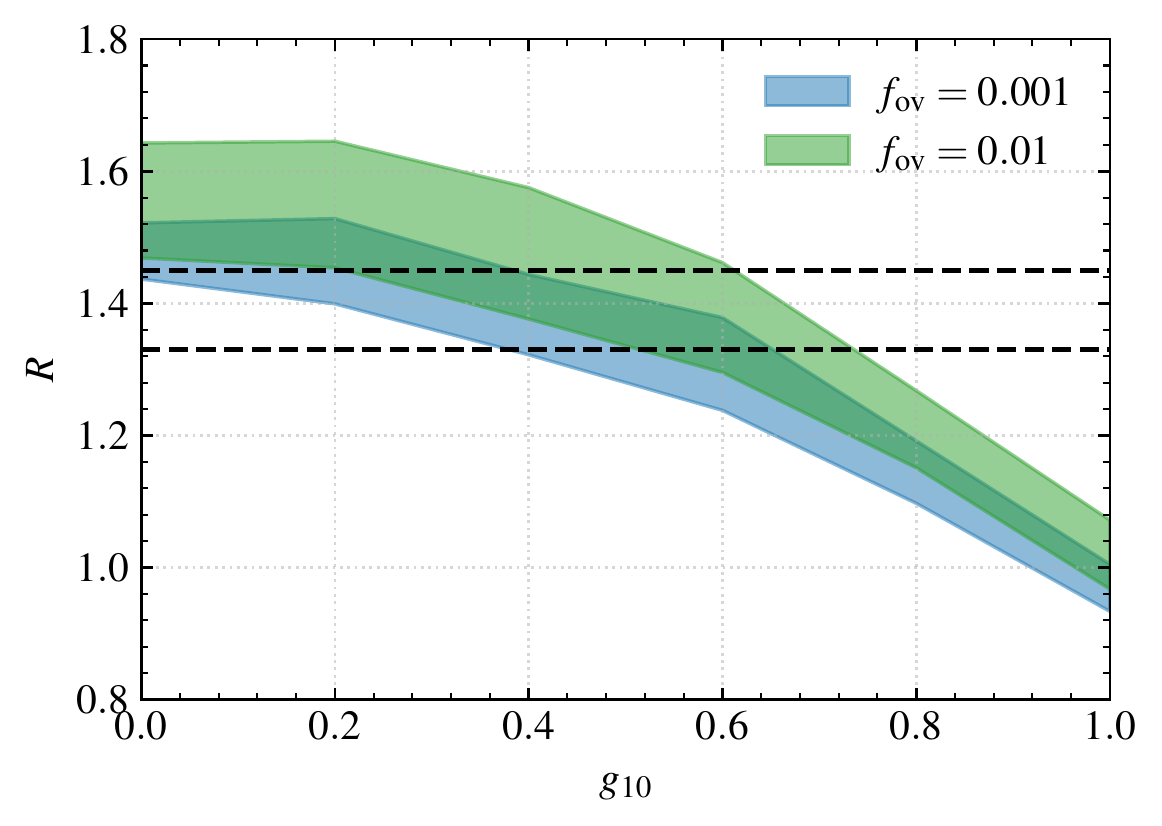}
     \includegraphics[width=0.47\textwidth]{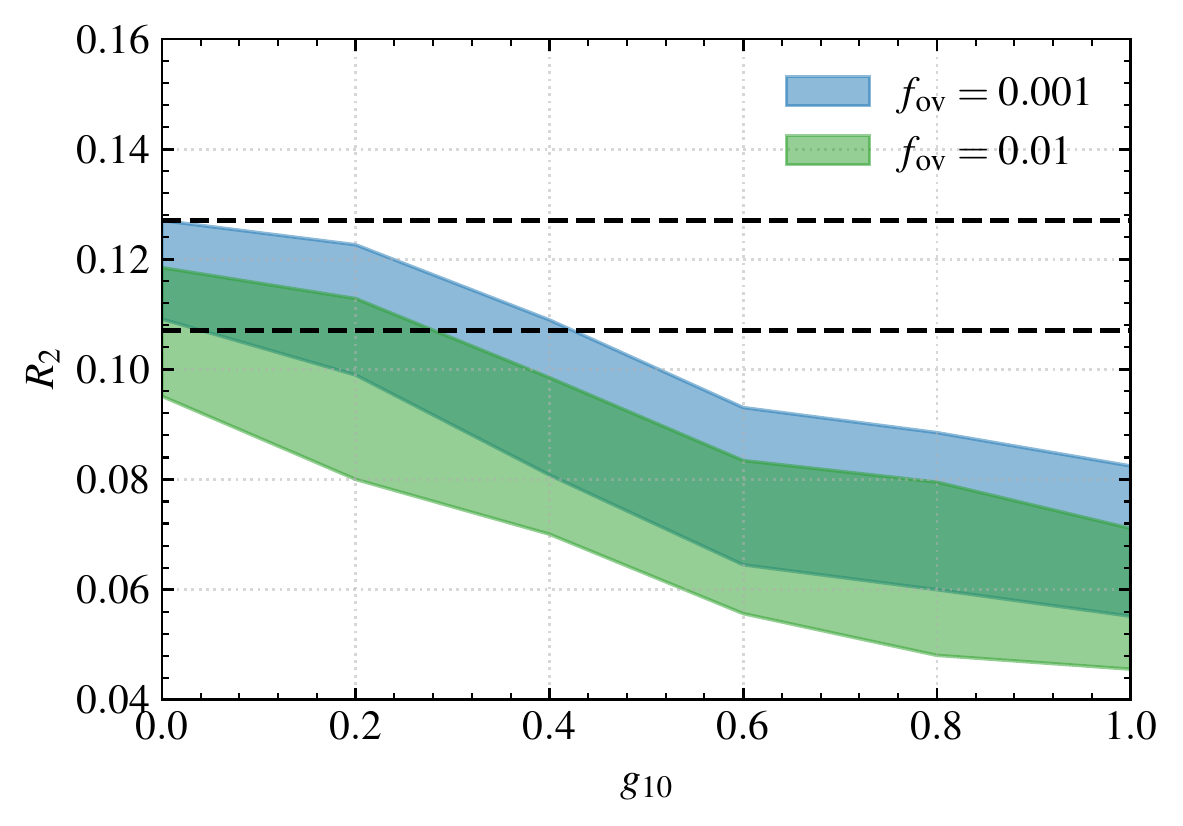}
    \caption{(Left panel): Predicted values of $R$ as a function of $g_{10}$ given standard convective core overshoot with $f_{\mathrm{ov}}=0.001$ (blue) and $f_{\mathrm{ov}}=0.01$ (green). The observed limit on $R$ is indicated by the region between the dashed black lines (95\% C.I.). (Right panel): The full range of $R_2$ values predicted as functions of $g_{10}$ given standard overshoot with $f_{\mathrm{ov}}=0.001$ (blue) and $f_{\mathrm{ov}}=0.01$ (green). The observed limit is again shown by the dashed black lines. }
    \label{fig: R and R2}
\end{figure*}

\subsection{Convective boundaries}
\label{sec: CBs}

Mixing beyond convective boundaries is a fundamental issue in stellar astrophysics and, in particular, the modelling of the HB phase. For a review of this phenomenon, see Ref \cite{2017RSOS....470192S}.

The root of this issue is in the chemical evolution of the core. Helium burning, which comprises the triple-$\alpha$ process and $^{12}C(\alpha, \gamma)^{16}O$ reaction, enriches the core with carbon and oxygen. This new mixture is more opaque than that which it replaces, and therefore moves the core further from convective neutrality. Consequently, any movement of carbon/oxygen across the convective boundary, which is physically favoured, increases the opacity of the neighbouring stable region and causes the core to grow. Simultaneously, the influx of helium into the core associated with this mixing decreases the opacity therein, which ultimately results in the splitting of the core. Repeated episodes of growth and splitting seed the development of an intermediate region which frequently switches between stability and instability.


The most dramatic consequence of these uncertainties occurs during the late stages of HB evolution, when the supply of helium in the core is dwindling. In these circumstances, the injection of even a small amount of helium into the core can spark large convective episodes which penetrate deeply within the helium-rich region surrounding the core. These events, which are known as \textit{core breathing pulses} (CBPs), can cause the significant elongation of the HB phase. Whether CBPs are physical in origin or merely a numerical phenomenon is still a topic of debate in stellar astrophysics \cite{2017RSOS....470192S}.

Schemes for modelling the structural evolution of the convective core can vary both in the manner they model mixing at the convective boundary, and within the intermediate region defined above. We categorise those available in \texttt{MESA} into two broad approaches: overshoot-induced and instantaneous mixing.

\subsubsection{Overshoot-induced mixing}
\label{sec: overshoot-based mixing}

Convective elements arrive at the convective boundary (where their acceleration falls to zero) with non-zero momentum, and therefore penetrate some distance within the stable region. This phenomenon is known as convective overshoot.

The canonical approach to modelling convective overshoot is as a time-dependent diffusive process with diffusion coefficients that decrease exponentially with distance from the convective boundary \cite{Herwig1997}. The rate of this exponential decrease is set by a free-parameter $f_{\mathrm{ov}}$, which is typically calibrated to observation. A definition of this parameter is provided in Appendix \ref{app: overshoot-based approaches}. 

Two of the schemes we consider in our simulations are based on convective overshoot. These are:
\begin{enumerate}
    \item \textit{Standard overshoot} (SO), which employs convective overshoot without specifying how to model mixing in the intermediate region defined above.
    \item \textit{Semiconvection} (SC), which uses convective overshoot in conjunction with a form of slow \textit{semiconvective} mixing in the intermediate region.
\end{enumerate}

\subsubsection{Instantaneous mixing}
\label{sec: instantaneous mixing}

\texttt{MESA}'s instantaneous mixing schemes were introduced to address theoretical shortcomings with their initial approach to locating the convective boundary. They operate by allowing the core to grow if the  model cells surrounding some initial candidate also become convective should their contents be homogeneously mixed with the rest of the core. The algorithm they employ progresses cell by cell until one is found which does not become unstable even after this mixing. The previous candidate is then identified as the formal convective boundary. This mixing occurs during a single timestep and is therefore instantaneous.

There are two general implementations of this process:
\begin{enumerate}
    \item \textit{Predictive Mixing} (PM), which employs this approach without committing chemical changes during the interim steps to the stellar model until the entire routine has completed.
    \item \textit{Convective Premixing} (CP), which improves upon further theoretical shortcomings of PM by updating and saving changes in chemical abundances for each intermediate step in the algorithm. This ensures that the convective boundary actually coincides with a region of convective neutrality.
\end{enumerate}

The instantaneous nature of these schemes makes them particularly prone to core breathing pulses. These can be avoided in Predictive Mixing by limiting the rate at which helium is ingested into the core, following the \textit{Spruit overshoot} scheme of Refs \cite{Spruit2015} and \cite{Lattanzio3}. Predictive Mixing also enables the splitting of the core to be avoided, by slowing its growth when any region of the core gets too close to convective neutrality, mimicking the \textit{maximal overshoot} scheme of Refs \cite{Lattanzio2} and \cite{Lattanzio1}.  

Convective Premixing does not contain these options, and CBPs can only be avoided by either resisting increases in the helium mass fraction $Y$ within the convective region, or by switching to a different mixing scheme as the star approaches the TAHB.

In the next section we describe the manner in which $R$ and $R_2$ can be calculated from our simulations of globular cluster stars. For demonstrative purposes, these are presented for simulations employing standard overshoot only. However, all CB schemes identified in this section are considered in our final constraint in Section \ref{sec: results}.

\section{Globular cluster constraints}
\label{sec: GC constraint}

\subsection{Calculating globular cluster parameters}

\begin{figure*}[t]
    \centering
    \includegraphics[width=0.6\textwidth]{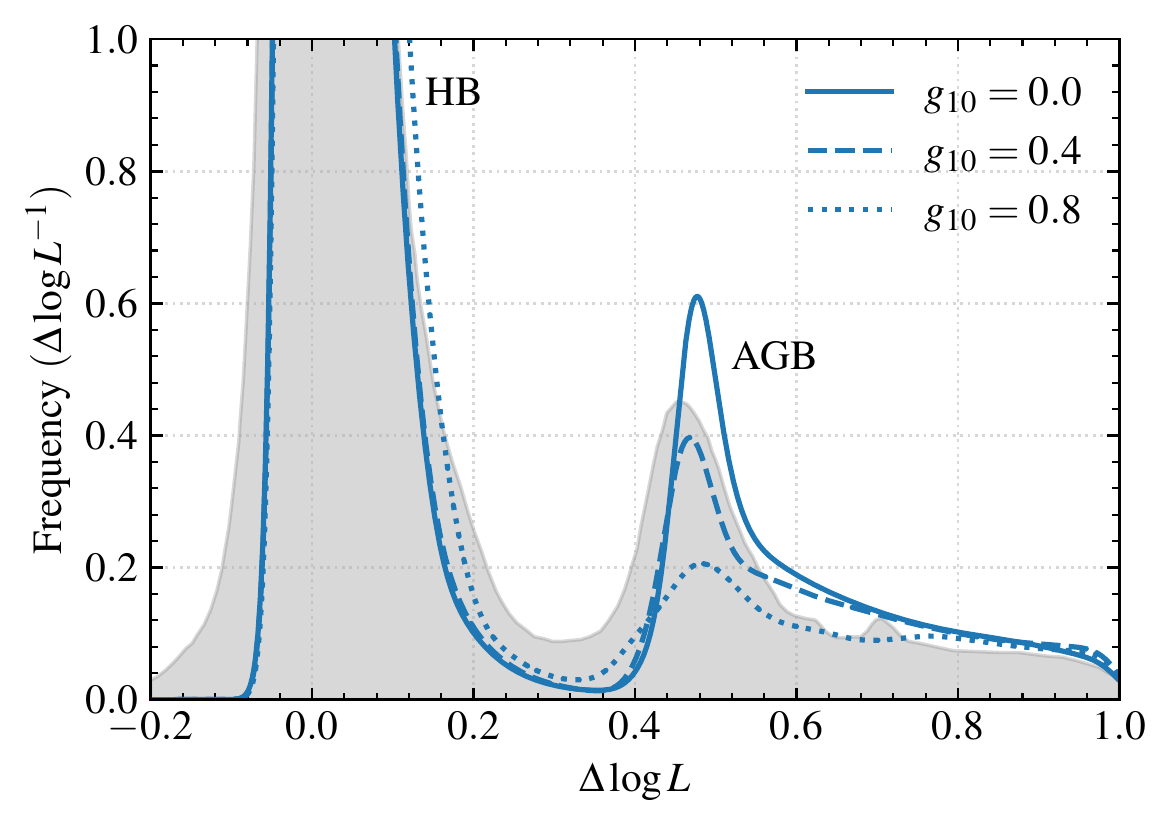}
    \caption{Theoretical $\Delta\log L$ PDFs obtained after averaging the 20 simulations given standard overshoot on the HB with $f_{\mathrm{ov}}=0.001$. The solid line corresponds to no axion energy-loss ($g_{10}=0.0$), while the dashed and dotted lines have experienced moderate ($g_{10}=0.4$) and strong ($g_{10}=0.8$) axion energy-loss. The solid grey region is the observed $\Delta\log L$ distribution presented in \cite{Lattanzio2}. The HB and AGB peaks are labelled. } 
    \label{fig: L PDF}
\end{figure*}

\subsubsection{The \texorpdfstring{$R$}{R}-parameter}
\label{sec: R-param}
If both the red giant and horizontal branches are well-populated, we can calculate $R$ as \cite{Raffelt:1996wa} 
\begin{equation}
    \label{eq: R}
    R_{\mathrm{th}}=\frac{\tau_{\mathrm{HB}}}{\tau_{\mathrm{RGB}}},
\end{equation}
where $\tau_{\mathrm{HB}}$ and $\tau_{\mathrm{RGB}}$ are the HB and RGB lifetimes for a single stellar model\footnote{Typically a set of stellar isochrones is necessary to determine star cluster parameters such as $R$. In the case of the horizontal and red giant branches of globular clusters, which have almost identical masses and ages (compared with the total age of the cluster), single stellar tracks are essentially isochrones themselves.}.

Calculating $R$ is therefore as simple as computing $\tau_{\mathrm{HB}}$ and $\tau_{\mathrm{RGB}}$ for each stellar model. Further details of this process, including the relevance of RGB stellar mass-loss, can be found in Appendix \ref{app: R-param}.

Using \texttt{MESA}, we have calculated $R$ for values of $g_{10}$ between $0.0$ and $1.0$ in intervals of $0.2$ given $f_{\mathrm{ov}}=0.001$ and $f_{\mathrm{ov}}=0.01$. For each unique set of input physics, 20 HB and AGB simulations were computed with different spatial and temporal resolutions to sample the sizeable stochastic variation present. \update{}

These sets of results are shown in the left panel Fig. \ref{fig: R and R2}, where the blue and green regions correspond to simulations with $f_{\mathrm{ov}}=0.001$ and $f_{\mathrm{ov}}=0.01$ respectively. The observed 95\% confidence limits on $R$ from Ref \cite{Ayala:2014pea} are indicated by the dashed black lines.

The presence of CBPs of varying number and duration within these sets of simulations results in a large spread in our predictions for $R$. Crucially, these CBPs can only extend HB lifetimes and increase predicted $R$-values. Consequently, the lower boundaries in the left-panel of Fig.~\ref{fig: R and R2} correspond to simulations with no CBPs, while those which define the upper boundaries have multiple long pulses. While these distributions are typically peaked towards the lower end, multiple simulations in our sets of 20 tend to be near each boundary.

In principle, we could repeat these simulations a large number of times and identify a theoretical 95\% confidence interval for these data. However, this proved computationally impractical and, consequently, we define our constraint when the entire predicted range from the 20 simulations performed falls below the \textit{observed} 95\% confidence interval~\footnote{It should be noted that this procedure likely introduces some uncertainty due to under-sampling into these values of $g_{10}^R$. For physical reasons these same issues do not affect bounds derived from $R_2$ (see Section~\ref{sec: R2parameter}), and hence will not influence our final results.}. For these example cases, such limits would occur at $g_{10}^R=0.65$ and $g_{10}^R=0.74$ for $f_{\mathrm{ov}}=0.001$ and $0.01$ respectively. Clearly, in this example scenario, increasing the overshoot parameter (which promotes greater mixing across convective boundaries) shifts the predictions for $\tau_{\mathrm{HB}}$, and therefore $R$, upwards. The magnitude of variation in the spread of $R$ is also larger in this case.

We also note that both sets of simulations show some preference for non-zero values of $g_{10}$. This horizontal branch hint is one of several well-known astrophysical anomalies which can be improved by the addition of novel energy-loss~\cite{Giannotti2017_recipes, DiLuzio:2021ysg}. While a thorough analysis of this hint is beyond the scope of our paper, we do note that the best-fit values of $g_{10}$ are sensitive to the adopted value of $Y_{\mathrm{init}}$. The simulations presented in Fig. \ref{fig: R and R2} employ $Y_{\mathrm{init}}=0.261$, the upper limit indicated by observational constraints \cite{Aver2013}, which most strongly emphasises its presence.

\subsubsection{The \texorpdfstring{$R_2$}{R2} parameter}
\label{sec: R2parameter}

To calculate $R_2$ we use the method outlined in Ref~\cite{Lattanzio2}, which, for clarity, is summarised here.

First, we convert the evolutionary tracks determined using \texttt{MESA} to probability density functions of $\Delta \log L=\log L-\log L_{\mathrm{HB}}$, the (log) luminosity difference from the HB peak. We do this using Equation 7 of Ref~\cite{Lattanzio2},
\begin{equation}
    P(\Delta\log L) = \frac{1}{\tau}\sum_{i=1}^n\frac{\Delta t_i}{\sigma\sqrt{2\pi}}\exp\bigg(-\frac{(\Delta\log L-\Delta\log L_i)^2}{2\sigma^2}\bigg),
\end{equation}
where $n$ is the number of timesteps between the ZAHB and the point where $\Delta\log L = 1.0$, $i$ is the model number, $\delta t_i$ is the elapsed time between models $i$ and $i+1$ and $\tau$ is the total time elapsed between $i=1$ and $i=n$. The value $\sigma=0.02$ was chosen to guarantee smooth theoretical PDFs, as in~\cite{Lattanzio2}.

Examples of these PDFs are provided in Fig. \ref{fig: L PDF} for $f_{\mathrm{ov}}=0.001$, given $g_{10}=0.0$ (solid), $g_{10}=0.4$ (dashed) and $g_{10}=0.8$ (dotted). The observed distribution from Globular Clusters~\cite{Lattanzio2} is indicated by the solid grey region. The maximum around $\Delta \log L=0$, which peaks at $\sim 14$, corresponds to the HB clump, while the AGB clump is given by the considerably smaller peak between $\Delta \log L=0.4$ and $\Delta \log=0.6$.

The distributions in Fig.~\ref{fig: L PDF} have a clear minimum around $\Delta \log L=0.3$. We take $R_2$ as the ratio of the area under the relevant curve to the right of the minimum (corresponding to the AGB peak) over the area to the left of the minimum (corresponding to the HB peak). This mimics the method for determining $R_2$ empirically from star cluster counts \cite{Lattanzio2}.

The right panel of Fig. \ref{fig: R and R2} shows example predictions for $R_2$ as a function of $g_{10}$ for the same set of simulations considered in Section \ref{sec: R-param}. The blue and green regions again indicate simulations with $f_{\mathrm{ov}}=0.001$ and $f_{\mathrm{ov}}=0.01$ respectively, and the observed 95\% confidence limits are given by the dashed black lines. The finite width of these regions is again due to the presence of CBPs within HB simulations. However, unlike with $R$, the stellar tracks with no CBPs lead to the highest predictions for $R_2$. As several such tracks appear in our set of 20 HB and AGB simulations, we can confidently define our constraint when the full range of predictions falls below the observed 95\% confidence limit. This occurs at $g_{10}^{R_2}=0.43$ when $f_{\mathrm{ov}}=0.001$ and $g_{10}^{R_2}=0.28$ when $f_{\mathrm{ov}}=0.01$.

The reciprocal dependence of $R_2$ on $\tau_{\mathrm{HB}}$ is evident in the right panel of Fig. \ref{fig: R and R2}. As previously discussed, larger values of $f_{\mathrm{ov}}$ (i.e. more efficient mixing across boundaries) increases the duration of the HB, thereby decreasing predictions of $R_2$. This effect is compounded, as larger overall helium consumption during the HB is known to lead to shorter AGB lifetimes \cite{Lattanzio2}. Importantly, this means that the free parameters associated with convective boundary mixing schemes have the opposite effects on $R$ and $R_2$, while energy-loss to axions reduces them both.

Finally, we note that the $R_2$ parameter does not show any preference for non-zero values of $g_{10}$ and thus does not contribute to the axion stellar energy loss "hints" mentioned in the previous sub-section. These predictions are consistent across different values of $Y_{\mathrm{init}}$, emphasising the robustness of $R_2$ against changes in chemical composition.

\subsection{Results}
\label{sec: results}

To derive our constraint, we performed the simulations described in Section \ref{sec: simulations} using a number of different implementations of standard overshoot, semiconvection, predictive mixing and convective premixing. Predictions of $R$ and $R_2$ for each of these were then generated along with their implied limits on the axion-photon coupling, $g_{10}^R$ and $g_{10}^{R_2}$. Details of the adopted model parameters for each scheme are discussed in Appendix \ref{sec: CB modelling}.

To account for the degeneracy with initial helium content, simulations were performed with identical input physics, but with $Y_{\mathrm{init}}=0.246$, $0.254$ and $0.261$, which correspond to the lower, central and upper 95\% confidence limits from low metallicity H II regions \cite{Aver2013}, the same constraint employed in Ref \cite{Ayala:2014pea}. Consequently, we have three values of $g_{10}^R$ and $g_{10}^{R_2}$ for each unique specification of convective boundary scheme. We then compare the most conservative values of $g_{10}^R$ and $g_{10}^{R_2}$ and select the most restrictive of these two as the constraint from globular clusters for that set of input physics. These values can then be compared across different CB schemes, with the least restrictive taken as the final constraint.

This results in overall constraints for each scheme of $g_{10}=0.43$ (SO), $g_{10}=0.47$ (SC), $g_{10}=0.28$ (PM) and $g_{10}=0.37$ (CP). The most conservative of these, which is obtained when semiconvection is used to model mixing across convective boundaries during the HB, is shown in deep violet in the ALP-plane alongside other relevant constraints in Fig. \ref{fig: param space}.
Even this constraint improves upon the pre-existing $R$-parameter constraint by approximately 30\%. A full table of results for these schemes and their model parameters is included in Appendix \ref{app: Results}.

Interestingly, for all of these schemes, $R_2$ is the more restrictive of the two globular cluster parameters considered. This is largely due to the impact of the initial helium content, as $g_{10}^R$ only falls below $g_{10}^{R_2}$ for the lowest considered value of $Y_{\mathrm{init}}$. 

Furthermore, while one might expect there to exist some choice of convective boundary model parameters which causes $g_{10}^R$ to fall below $g_{10}^{R_2}$, this does not seem to be the case. For example, in Section \ref{sec: sec2} we showed that decreasing $f_{\mathrm{ov}}$ from $0.01$ to $0.001$ reduced $g_{10}^R$ and increased $g_{10}^{R_2}$. However, the further reduction of $f_{\mathrm{ov}}$ by another factor of 10 does not have a major impact on the obtained values of these constraints.

Given the constraints presented in this section represent the most conservative values obtained for each CB scheme, it is likely that these can be refined as the uncertainties surrounding the modelling of these phenomena decrease. For example, complementary evidence from asteroseismology suggests that the predicted asymptotic g-mode $l=1$ period spacing in HB stars, which depends sensitively on the adopted convective boundary mixing scheme, is systematically lower than is indicated by observation \cite{Lattanzio1}. Of the CB schemes considered in this work, only predictive mixing goes some way to reconciling this tension. However, as the magnitude of the g-mode period spacing is dependent on possible mode trapping in semiconvective or partially-mixed zones, which can significantly increase the inferred period spacing \cite{Lattanzio1}, we choose not to adopt this more restrictive bound as our final constraint. It is nevertheless included in pink in Fig. \ref{fig: param space} for comparison.

\subsection{Other systematic uncertainties}
\label{sec: systematics}

\begin{table*}
\centering
\begin{tabular}{>{\centering}m{0.25\textwidth}>{\centering}m{0.06\textwidth}>{\centering}m{0.06\textwidth}>{\centering\arraybackslash}m{0.06\textwidth}} 
\hline\hline
Convective Boundary & \multicolumn{3}{c}{$^{12}C(\alpha, \gamma)^{16}O$} strength \\
Scheme   & 0.8   & 1.0   & 1.2                                 \\ 
\hline
Standard Overshoot                        & 0.45 &  0.43 & 0.33                               \\
Semiconvection                        & 0.31 &  0.47 & 0.37                               \\
Predictive Mixing                       & 0.34 &  0.28 & 0.13                               \\
Convective Premixing                          & 0.37 &  0.37 & 0.26                               \\
\hline\hline
\end{tabular}
\caption{This table shows the variation in the limit on the axion-photon $g_{a\gamma\gamma}$ as the $^{12}C(\alpha, \gamma)^{16}O$ is varied within its stated uncertainties, for the four convective boundary schemes we consider. \label{tab: RR table}}
\end{table*}

Given our new constraint comes from observed limits on the $R_2$-parameter, it is vital to assess the impact of any other sources of systematic uncertainty. These were discussed comprehensively in Ref \cite{Lattanzio2}, which predominantly identified the $^{12}C(\alpha,\gamma)^{16}O$ reaction, which is relevant in the latter stages of central helium burning, as the most important theoretical uncertainty affecting $R_2$ apart from convective boundary modelling. Mass loss during the RGB phase is also a source of uncertainty, and we adopt the most conservative implementation for this.

Our simulations employ the $^{12}C(\alpha,\gamma)^{16}O$ reaction rate from the NACRE II collaboration, which is accurate to within 20\% of its stated value in the temperature range of interest \cite{2013NuPhA.918...61X}. To account for the impact of this uncertainty, we repeated our simulations with $^{12}C(\alpha,\gamma)^{16}O$ multiplied by constant factors of $0.8$ and $1.2$. The results of this are shown in Table \ref{tab: RR table}.

For all schemes increasing the efficiency of this reaction rate leads to more restrictive constraints, though the magnitude of this effect varies between schemes. When the reaction rate is reduced, the obtained values of $g_{10}^{R_2}$ increase for the SO, PM and CP schemes, though the magnitudes of these shifts are not large.

This level of variation indicates that the uncertainty from $^{12}C(\alpha,\gamma)^{16}O$ in this work is smaller than that in Ref \cite{Lattanzio2}. This is primarily because their simulations employ the original NACRE $^{12}C(\alpha,\gamma)^{16}O$ reaction rate, which has upper and lower limits that are within 40\% of the central values \cite{ANGULO19993}.

\section{Conclusion}
\label{sec: conclusion}

We improve upon the $R$-parameter constraint on axions and axion-like particles by utilising another globular cluster parameter, the ratio of asymptotic giant branch to horizontal branch stars (the $R_2$-parameter). This approach has been made possible by the development in Ref \cite{Lattanzio2} of an observed limit on $R_2$ which is more statistically robust and self-consistent than those of previous analyses.

We showed, using the stellar evolution code Modules for Experiments in Stellar Astrophysics (\texttt{MESA}),  that the addition of axion energy-loss via the Primakoff effect - which is more efficient in the hotter helium burning shells of AGB stars than in HB cores - reduces the predicted value of $R_2$.

The value of our constraint on $g_{a\gamma\gamma}$ depends sensitively on the employed method for modelling mixing across convective boundaries during the horizontal branch phase. While the issues surrounding convective boundaries are well-documented in astrophysical literature, their effects have not previously been quantified in beyond the Standard Model constraints. Four different schemes available in \texttt{MESA} were employed to quantify this spread: standard overshoot, semiconvection, predictive mixing and convective premixing.   Important sources of systematic uncertainty, including the initial helium content of the globular clusters, the mass-loss rates during the RGB phase and the $^{12}C(\alpha,\gamma)^{16}O$ reaction rate were taken into account. While the last of these was expected to significantly affect our results, improvement in the accuracy of this nuclear reaction rate has drastically reduced its impact. The most conservative constraint was obtained using semiconvection, yielding $g_{a\gamma\gamma} \leq 0.47\times10^{-10}$ GeV$^{-1}$.

Evidence from asteroseismology favours the use of predictive mixing, which supports larger convective cores. Were this to be shown conclusively in future work, our constraint would be  $g_{a\gamma\gamma} \leq 0.34\times10^{-10}$ GeV$^{-1}$. A greater understanding of stellar evolution will come from analyses using GAIA and James Webb Space Telescope data, leading to even stronger bounds.
Some of the region of parameter space we study will be further probed by axion haloscopes and other proposals such as the IAXO family of experiments~\cite{IAXO:2020wwp} and ALPS-II~\cite{Bahre:2013ywa}, although our bound extends to higher axion masses than those experiments can attain. Searches for axions will continue to be an exciting and active part of the particle physics landscape in the years to come.

\appendix

\section{Axion energy-loss}
\label{app: axion energy-loss}

Axions with masses less than a few keV and which interact primarily via Equation \ref{eq: axion-photon L} are produced in stellar environments by the Primakoff process ($Ze+\gamma\rightarrow Ze+a$). To accurately model axion energy-loss in all evolutionary phases considered, we employ the methods of Ref \cite{Raffelt:1987yu}.

The relevant form of the energy-loss rate per unit mass to Primakoff production in a non-degenerate medium is given by \cite{Raffelt:1987yu}
\begin{equation}
    \label{eq: Primakoff energy loss}
    \epsilon_{\mathrm{nd}}=\frac{g_{a\gamma\gamma}^2T^7}{16\pi^2\rho}y_1^2F(y_0, y_1),
\end{equation}
where $\rho$ is the stellar mass density. The function $F(y_0,y_1)$ takes the form
\begin{equation}
    \label{eq: F function}
    F(y_0, y_1)=\frac{1}{4\pi}\int_{y_0}^{\infty}dy\frac{y^2(y^2-y_0^2)^{1/2}}{e^y-1}I(y,y_0,y_1),
\end{equation}
where
\begin{equation}
    I=\int_{-1}^{+1}dx\frac{1-x^2}{(r-x)(r+s-x)}
\end{equation}
and
\begin{equation}
    r=\frac{2y^2-y_0^2}{2y(y^2-y_0^2)^{1/2}},\ s=\frac{y_1^2}{2y(y^2-y_0^2)^{1/2}}.
\end{equation}

In these equations $y_1=k_{\mathrm{nd}}/T$ is the temperature normalised Debye-H\"{u}ckel wavenumber, which sets the scale of screening effects and is given by
\begin{equation}
    \label{eq: DH wavenumber}
    k_{\mathrm{nd}}^2=\frac{4\pi\alpha}{T}n_B(Y_e + \sum_j Z_j^2Y_j),
\end{equation}
where $\alpha$ is the electromagnetic fine structure constant, $n_B$ is the baryon number density, $Z_j$ is the charge of nuclear species j and $Y_j$ ($Y_e$) is the number fraction per baryon of species $j$ (electrons): $Y_{j,e}=n_{j,e}/n_B$ where $n$ is number density.

The parameter $y_0$ in Equation \ref{eq: F function} is the temperature normalised plasma frequency $y_0=\omega_{\mathrm{pl}}/T$, with $\omega_{\mathrm{pl}}=4\pi\alpha n_e/m_e$. If stellar evolution on the horizontal branch alone was being considered, the limit $\omega_{\mathrm{pl}}\approx0$ could be taken without great consequence. However, this is not valid in RGB stars where $\omega_{\mathrm{pl}}$ can exceed the temperature $T$.

To accurately model energy-loss during the RGB and AGB phases, the total energy-loss must be separated into electron and ion components, i.e.
\begin{equation}
    \epsilon_{\mathrm{d}}=\epsilon_{\mathrm{ions}} + \epsilon_{\mathrm{e}}.
\end{equation}
The energy-loss to ions can be calculated as
\begin{equation}
    \label{eq: ion energy loss}
    \epsilon_{\mathrm{ions}}=\frac{g_{a\gamma\gamma}^2T^7}{16\pi^2\rho}y_2^2F(y_0, y_2),
\end{equation}
where now $y_2=k_{\mathrm{ions}}/T$, with
\begin{equation}
    k_{\mathrm{ions}}^2 = \frac{4\pi\alpha}{T}\sum_{\mathrm{ions}}Z_j^2n_j .
\end{equation}

The contribution of degenerate electrons is given by \cite{Raffelt:1987yu}
\begin{equation}
    \label{eq: electron energy loss}
    \epsilon_e=\frac{\alpha g_{a\gamma\gamma}^2}{4\pi}\frac{T^4}{\rho}R_{\mathrm{deg}}n_eF(y_0, y_3),
\end{equation}
where $R_{\mathrm{deg}}$ is the Pauli-reduction factor and the screening scale is now set by $y_3=k_{\mathrm{TF}}/T$, with $k_{\mathrm{TF}}$ the Thomas-Fermi wavenumber
\begin{equation}
    k_{\mathrm{TF}}^2=\frac{4\alpha m_ep_F}{\pi}.
\end{equation}

The final energy-loss rate per unit mass is then calculated in each cell as a linear combination of the degenerate and non-degenerate rates
\begin{equation}
    \epsilon_a=(1-w)\epsilon_{\mathrm{nd}} + w(\epsilon_{\mathrm{ions}}+\epsilon_e).
\end{equation}
The parameter $w$ is given as a function of $\zeta=(3\pi^2n_e)^{2/3}/2m_eT$ as
\begin{equation}
    w=\frac{1}{\pi}\arctan(\zeta-3)+\frac{1}{2}.
\end{equation}
The Pauli-reduction factor $R_{\mathrm{deg}}$, which appears in Equation \ref{eq: electron energy loss}, is given by
\begin{equation}
    \frac{1.5}{\max(1.5, \zeta)}.
\end{equation}

We included these equations in \texttt{MESA} by adding terms to the module \texttt{neu}, which is responsible for calculating the energy-loss associated with neutrino production. This approach is based on Ref \cite{Friedland} and its corresponding \texttt{MESA} test-suite case. Our code is available at the following link\footnote{\label{fn: link}\url{https://github.com/fhiskens/Globular_Clusters}}. As evaluating the function $F$ exactly within the code is impractical, we follow Ref \cite{Raffelt:1987yu} and calculate it by interpolating between pre-processed the slowly varying function $G(y_0, y_1)$ where 
\begin{equation}
    F(y_0, y_1)=\frac{100}{1+y_1^2}\frac{1+y_0^2}{1+e^{y_0}}G(y_0, y_1).
\end{equation}

\section{Input physics}
\label{app: simulations}

In this section we describe the input physics used for our simulations. We provide our inlists and \texttt{run\_star\_extras} files at the link in Footnote \ref{fn: link}.

Much of our input physics follows the choices of the \texttt{MESA} Isochrones and Stellar Tracks (MIST) group. These include our choice of opacity tables, atmospheric boundary conditions, convective envelope overshoot, element diffusion (which is only turned on during main sequence evolution), AGB stellar wind scheme, mixing-length prescription and nuclear reaction rates, the details of which can be found in the original MIST papers \cite{MIST_0, MIST_1}. 

Instead of employing the full \texttt{mesa\_49.net} network favoured by MIST, we use the smaller \texttt{pp\_cno\_extras\_o18\_ne22.net} to reduce the computational burden of our simulations. This choice still allows for in-depth realisations of the proton-proton chain and CNO cycle during hydrogen burning. Given the mass range being considered we need not worry about details concerning carbon burning and beyond. This input physics is kept the same for all simulations performed throughout this work. For the most part we adopt a metallicity of $Z=0.001$ and initial mass $M_{\mathrm{init}}=0.82$. The adopted initial helium abundance $Y_{\mathrm{init}}$ is discussed in Section \ref{sec: results}.

In addition to the input physics specified above, we include mass loss following the Reimers equation \cite{Reimers}
\begin{equation}
    \label{eq: Reimers}
    \Dot{M}_\mathrm{R}=4\times10^{-13}\eta_{\mathrm{R}}\frac{(L/L_{\odot})(R/R_{\odot})}{(M/M_{\odot})}\ M_{\odot}\ \mathrm{yr}^{-1},
\end{equation}
during RGB evolution, where $L$, $R$ and $M$ are the stellar luminosity, radius and mass respectively and $\eta_{\mathrm{R}}$ is a dimensionless parameter of order unity. As shall be discussed in Appendix \ref{app: R-param}, multiple values of $\eta_{\mathrm{R}}$ are considered for our pre-HB simulations.

\section{The modelling of convective boundaries}
\label{sec: CB modelling}
The flagship systematic uncertainties under consideration are those associated with the modelling of mixing across convective boundaries during the horizontal branch evolutionary phase. Here we provide a more in-depth description of the issues this presents, as well as the various approaches to modelling this phenomenon included in \texttt{MESA}.

\subsection{The formal convective boundary}

In stellar modelling, the stability of a region against convection is indicated by the relative magnitudes of the \textit{radiative} and \textit{adiabatic} temperature gradients. The former is defined as \cite{2012sse..book.....K}
\begin{equation}
    \nabla_{\mathrm{rad}}=\bigg(\frac{\partial \ln T}{\partial \ln P}\bigg)_{\mathrm{rad}}=\frac{3}{16\pi a_R G}\frac{\kappa l P}{mT^4},
\end{equation}
and represents the temperature gradient which would occur in a stellar region if radiation alone were responsible for energy transport. Here $a_R$ and $G$ are the radiation density and gravitational constants respectively and $\kappa$, $l$, $P$, $m$ and $T$ are the opacity, luminosity flux, pressure, enclosed mass and temperature at radial location $r$. The adiabatic temperature gradient, given by
\begin{equation}
    \nabla_{\mathrm{ad}}=\bigg(\frac{\partial \ln T}{\partial \ln P}\bigg)_{\mathrm{ad}},
\end{equation}
encodes the temperature gradient which would occur if energy transfer occurs via the movement of convective elements, which are assumed to behave adiabatically. This is typically calculated in stellar models using the mixing length theory (MLT) of convection \cite{2012sse..book.....K}.

According to the \textit{Schwarzschild} criterion \cite{Schwarzschild1906}, a region will be stable against convection if
\begin{equation}
    \label{eq: SC criterion}
    \nabla_{\mathrm{rad}}<\nabla_{\mathrm{ad}}.
\end{equation}

An alternative to the Schwarzschild criterion is the \textit{Ledoux} criterion for stability
\begin{equation}
    \label{eq: Ledoux}
    \nabla_{\mathrm{rad}}<\nabla_{\mathrm{ad}}+\frac{\varphi}{\delta}\nabla_{\mu}\equiv \nabla_{\mathrm{L}},
\end{equation}
which incorporates the stabilising effect of the mean molecular weight ($\mu$) gradient $\nabla_{\mathrm{\mu}}$, where
\begin{equation}
    \nabla_{\mu}=\frac{d\ \ln \mu}{d\ \ln P},\ \varphi=\bigg(\frac{\partial\ln\rho}{\partial\ln\mu}\bigg)_{T,P}\, \mathrm{and}\,\, \delta=\bigg(\frac{\partial\ln\rho}{\partial\ln T}\bigg)_{P, \mu}. 
\end{equation}
The formal convective boundary is defined as the point of convective neutrality, i.e. $\nabla_{\mathrm{rad}}=\nabla_{\mathrm{ad}}$ for Schwarzschild or $\nabla_{\mathrm{rad}}=\nabla_{\mathrm{L}}$ for Ledoux, where the acceleration of convective elements falls to zero.

Such elements may arrive at the boundary with non-zero momentum and, consequently, penetrate within the radiative zone. Such a processes is termed \textit{convective overshoot}, and elicits chemical mixing across convective boundaries.

\subsection{CHeB convective boundaries}
\label{sec: CHeB CBs}

As alluded to in Section \ref{sec: CBs}, the uncertainty associated with modelling mixing beyond convective boundaries arises due to the chemical evolution of the core. During the HB phase, the triple-$\alpha$ process and  $^{12}C(\alpha, \gamma)^{16}O$ reaction respectively produce carbon (C) and oxygen (O), which is efficiently mixed throughout the whole convective core. This new mixture causes the core to have a higher opacity $\kappa$, and hence higher $\nabla_{\mathrm{rad}}$, than the helium rich mixture immediately outside it.

If convective overshoot has not been included in the simulation, a large discontinuity in $\nabla_{\mathrm{rad}}$ develops at the formal boundary, which remains stable and does not grow. On the other hand, if overshoot is included, as is suggested by theory and observation \cite{Lattanzio1, Lattanzio2, Lattanzio3}, the CO-rich mixture is transported into the region immediately outside the formal boundary. This again increases $\nabla_{\mathrm{rad}}$, making the region unstable to convection.

A direct consequence of the core growing to incorporate this formally stable region is the influx of additional helium into the convective zone. This reduces the opacity across the core, causing the ratio $\nabla_{\mathrm{rad}}/\nabla_{\mathrm{ad}}$ to approach 1 (i.e. convective neutrality). 

Curiously, $\nabla_{\mathrm{rad}}/\nabla_{\mathrm{ad}}$ profiles within the cores of CHeB stars can have minima which are displaced from the formal boundary. Consequently, the stellar region at which this minimum occurs is the first to reach convective neutrality, leading to the splitting of the core. This process is shown explicitly in Section 4 of Ref \cite{2017RSOS....470192S}.

HB cores can host numerous episodes of growth and splitting throughout their evolution, which typically results in the establishment of a \textit{semiconvective} region between the convective core-proper and the He-rich region above. Though purpose-built schemes exist to model semiconvection, similar results can be achieved without these should enough of these episodes of growing and splitting occur. While the former of these will produce a smooth chemical profile, the latter will be composed of (possibly many) discrete steps, depending on the number of episodes that occurred.

\subsection{Methods for modelling convective boundaries}
\texttt{MESA v12778} offers a number of different approaches to model mixing beyond convective boundaries and within the intermediate region. Of these we distinguish two broad categories: overshoot-based approaches and instantaneous mixing schemes. Note that it is possible to implement no additional mixing across convective boundaries. However, this results in the ratio $\nabla_{\mathrm{rad}}/\nabla_{\mathrm{ad}}>1$ on the convective side of the boundary, which is unphysical \cite{2017RSOS....470192S}. Quite apart from this, such a scheme results in predictions which are incompatible with observation. Consequently, we do not consider such an approach in this work.

\subsubsection{Overshoot-based approaches}
\label{app: overshoot-based approaches}

\texttt{MESA} treats convective overshoot as a diffusive process, with diffusion coefficient $D_{\mathrm{ov}}$ given by \cite{MESA1}
\begin{equation}
    \label{eq: overshoot diffusion}
    D_{\mathrm{ov}}=D_{\mathrm{conv},0}\exp\bigg(\frac{-2z}{f\lambda_{P,0}}\bigg),
\end{equation}
where $D_{\mathrm{conv},0}$ is the diffusion coefficient derived from mixing-length theory some non-zero distance within the convective region, $\lambda_{P,0}$ is the pressure scale height at this point (i.e. the distance over which pressure falls by a factor of $e$), $z$ is the radial distance from this point and $f$ is a free overshoot parameter. In \texttt{MESA}, the location of this reference zone, at which $D_{\mathrm{conv},0}$ and $\lambda_{P,0}$ are defined, is specified by the user to be a fraction of the pressure scale height \textit{within} the formal convective boundary. This fraction is set using the model parameter \texttt{overshoot\_f0}. The parameter $f_{\mathrm{ov}}$ used throughout this work indicates the fractional distance from the convective boundary (rather than the reference region) over which overshoot is desired. It is related to $f$ in Equation \ref{eq: overshoot diffusion} and \texttt{overshoot\_f0} as 
\begin{equation}
    f_{\mathrm{ov}}=f-\texttt{overshoot\_f0}.
\end{equation}
Typical values of $f_{\mathrm{ov}}$ are between $\mathcal{O}(0.001)$-$\mathcal{O}(0.01)$ (see e.g. Refs \cite{Lattanzio2} and \cite{MIST_1}). Larger values than this result in predictions of $R_2$ which are incompatible with observation.

The first of our overshoot-based approaches, which we term \textit{standard overshoot}, employs this form of convective overshoot across a convective boundary defined by the Schwarzschild criterion (Equation \ref{eq: SC criterion}). It does not include any specific semiconvective mixing, relying instead upon the repeated growth and splitting of the core to maintain the convective neutrality of the intermediate region.

Our second approach employs a direct form of \textit{semiconvection}. In \texttt{MESA}, semiconvective mixing is also implemented as a diffusive process, with diffusion coefficient $D_{\mathrm{sc}}$ calculated following the method of Ref \cite{Langer85},
\begin{equation}
    D_{\mathrm{sc}}=\alpha_{\mathrm{sc}}\bigg(\frac{K}{6C_P\rho}\bigg)\frac{\nabla_{\mathrm{rad}}-\nabla_{\mathrm{ad}}}{\nabla_{\mathrm{rad}}-\nabla_{\mathrm{L}}}.
\end{equation}
Here, $K$ is the radiative conductivity, $C_P$ is the specific heat at constant pressure and $\alpha_{\mathrm{sc}}$ is a free parameter which governs the strength of this form of mixing. Clearly larger values of $\alpha_{\mathrm{sc}}$ result in more efficient mixing. As noted in Ref \cite{MIST_1}, it is typical to see values of $\alpha_{\mathrm{sc}}$ between $0.001$ and $1.0$ in the literature.

This scheme is implemented in regions which are formally stable according to the Ledoux criterion, but unstable according to the Schwarzschild criterion. Convective overshoot is also included to induce mixing across the formal convective boundary, which is necessarily defined by the Ledoux criterion. Consequently there are two free parameters which must be defined in each implementation of this scheme - $f_{\mathrm{ov}}$ and $\alpha_{\mathrm{sc}}$.

\subsubsection{Instantaneous mixing schemes}
The typical approach to handling convective boundaries in \texttt{MESA} can lead to theoretical deficiencies in the models, a summary of which can be found in the \texttt{MESA} instrument papers \cite{MESA4, MESA5}.

In the context of the mixing length theory of convection, the condition that convective elements fall to zero (i.e. the definition of the boundary) is equivalent to the equality $\nabla_{\mathrm{rad}}=\nabla_{\mathrm{ad,\ L}}$. As convective velocities are only defined within the convective region, this condition must be met on the \textit{inside} of the convective boundary \cite{Gabriel2014}. Despite this, when combined with chemical discontinuities across convective boundaries, the overshoot-based schemes identified above produce scenarios where $\nabla_{\mathrm{rad}}>\nabla_{\mathrm{ad}}$ at the convective boundary.

An attempted solution to this issue, termed \textit{predictive mixing}, was introduced and documented in \texttt{MESA} instrument paper 4 \cite{MESA4}, based on the procedure of Ref \cite{Bossini2015} as well as the \textit{maximal overshoot} scheme of Ref \cite{Lattanzio1}.

The scheme begins by identifying a candidate convective boundary as the model cell for which one face is convective ($y\equiv\nabla_{\mathrm{rad}}/\nabla_{\mathrm{ad}}-1>0$) and the other is radiative ($y<0$). The change in $y$ at each respective face is then calculated again with the assumption that the cell's contents are homogeneously mixed with that of the neighbouring convective region. If, after this process, both faces have $y>0$, then the algorithm proceeds to consider the next nearest cell in the radiative region as its new candidate. This continues until a cell, whose radiative boundary does not become convective upon recalculation of $y$, is identified. The previous candidate is then identified as the formal convective boundary. No chemical abundance changes are incorporated into the model until the iterative process has finished.

Given the predictive mixing routine is executed for each timestep, it describes an instantaneous form of mixing. As argued in the instrument paper, this should not be an issue when modelling the HB, as the convective evolution is considerably faster than the characteristic timescale of helium burning \cite{MESA4}.

In addition to the core process presented above, predictive mixing is furnished with a number of other features to increase its applicability. The first of these, taking inspiration from the maximal overshoot scheme employed in Refs \cite{Lattanzio1, Lattanzio2, Lattanzio3}, aims to prevent the splitting of the core via the introduction of an additional free parameter \texttt{predictive\_superad\_thresh}, which we term $\delta_{\mathrm{PM}}$. If $y$ falls below the user-specified value of $\delta_{\mathrm{PM}}$ anywhere in the core, the routine reverts to the previous boundary candidate. Such a condition has been found to promote the maximal growth of the core and goes some way to reconciling differences between observed and predicted non-radial period spacing in HB asteroseismology \cite{Lattanzio1}.

As an instantaneous mixing scheme, predictive mixing is particularly prone to large and repeated core breathing pulses, which are disfavoured by observation. One way to prevent these is to limit the rate of ingestion of helium into the core from the surrounding radiative region \cite{Spruit2015}. This has been done following the implementation in Ref \cite{Lattanzio3}, which defines the maximum rate of helium ingestion as
\begin{equation}
    \dot{m}_i=\alpha_{i}\frac{L}{RT}\bigg(1-\frac{\nabla_{\mathrm{rad}}}{\nabla_{\mathrm{ad}}}\bigg),
\end{equation}
where $\alpha_{i}$ is the ingestion efficiency\footnote{Note that the ingestion efficiency as defined in \texttt{MESA} is related to that in the original paper as $\alpha_{i}=5\alpha_i^{\mathrm{Ref\ [57]}}/12$.} and $R$ is the gas constant. Smaller values of $\alpha_i$ hinder the influx of helium into the core and are more efficient at preventing CBPs. Values of $\delta_{\mathrm{PM}}=\mathcal{O}(0.005)$ are typical \cite{Lattanzio2}, while values of $\alpha_i=\mathcal{O}(0.1-1)$ are approximately consistent with observation in Ref \cite{Lattanzio3}.

The realisations of predictive mixing employed in this work therefore also have two free parameters -- $\delta_{\mathrm{PM}}$, for preventing the splitting of the core, and $\alpha_i$, which limits helium influx.

Despite the initial success of predictive mixing for preventing $\nabla_{\mathrm{rad}}>\nabla_{\mathrm{ad}}$ on the inside of the convective boundary in many stellar evolutionary phases, it could not do so for core helium burning stars. \texttt{MESA's} second instantaneous mixing scheme, \textit{convective premixing}, remedies this situation.

Convective premixing is applied at the beginning of each timestep, before structural and chemical changes have been computed. It employs an almost identical algorithm to the case of \textit{predictive mixing}. However, chemical changes induced by the addition of a candidate cell to the convective region are now committed to the model at each iteration.

A quirk of convective premixing is that newly transitioned cells are approximately convectively neutral. Consequently it can be viewed as a scheme for effective semiconvection. The finer points of this scheme and its implementation can be found in \texttt{MESA} Instrument Paper 5 \cite{MESA5}.

Like predictive mixing, convective premixing is also prone to CBPs owing to the instantaneous nature of its mixing. As of the current version of \texttt{MESA}, these can only be avoided by switching mixing scheme, or by telling the code to resist increases in the abundance of the species being burned (in our case helium). This is done by setting model parameter \texttt{conv\_premix\_avoid\_increase} to \texttt{.true.}. This is the only free parameter that appears in the realisations of the convective premixing scheme in this work.

\section{Further calculation details}
\label{sec: Calculation details}

In Section \ref{sec: GC constraint}, we provided the salient steps to calculate both $R$ and $R_2$ from  the evolutionary tracks. For the sake of brevity some details were omitted there and are instead provided here.

\subsection{The \texorpdfstring{$R$}{R}-parameter}
\label{app: R-param}

To determine $R$, we must calculate both the HB lifetime $\tau_{\mathrm{HB}}$ and the RGB lifetime $\tau_{\mathrm{RGB}}$. We define the former as the elapsed time between the ZAHB, when stable helium burning begins in the convective core, and the TAHB, when the central helium mass fraction falls below $10^{-4}$. This is easily extracted from our stellar evolution simulations.

The determination of the RGB lifetime, however, is slightly more involved. The observed values of $R$ used in Ref \cite{Ayala:2014pea} (which were originally presented in Ref \cite{Salaris2004}) only counted RGB stars with F555W magnitudes greater than that of the ZAHB. Our RGB lifetime must reflect this.

To replicate this choice theoretically, we calculated the F555W magnitude $V_{\mathrm{F555W}}(L, T_{\mathrm{eff}}, \log g)$ at every timestep in the evolutionary track using a suitable set of bolometric corrections (BCs). We use the HST Wide Field and Planetary Camera 2 (WFPC2) BC table provided on the MIST website\footnote{\url{https://waps.cfa.harvard.edu/MIST/model\_grids.html}}, computed from the C3K grid. $\tau_{\mathrm{RGB}}$ is then calculated as the elapsed time between the point on the RGB where $V_{\mathrm{F555W}}=V_{\mathrm{F555W}}^{\mathrm{ZAHB}}$ and the RGB tip, at which point helium ignition takes place.

An important consideration in the determination of $R$ is the adopted rate of mass loss during the RGB phase, which is typically modelled using Equation \ref{eq: Reimers}. The magnitude of the parameter $\eta_{\mathrm{R}}$ can have a sizeable impact upon the evolution of stars beyond the RGB. For example, a value of $\eta_{\mathrm{R}}=0.2$ results in only a minor decrease in the envelope mass by the ZAHB. If $\eta_{\mathrm{R}}=0.4$, however, a considerably larger fraction of the envelope will be lost throughout the RGB and the resulting HB stars will be \textit{hotter}, but similarly luminous \cite{2012sse..book.....K}.

These factors are relevant, as the $R$-parameter is canonically defined with a ZAHB effective temperature of $\log T_{\mathrm{eff}}=3.85$. As it would be impractical to individually tune each simulation to this ZAHB effective temperature, we instead follow the method of \cite{Ayala:2014pea} and run two simulations, one with $\eta_{\mathrm{R}}=0.2$ and another for which $\eta_{\mathrm{R}}=0.4$, calculate $R$ for both of these cases and interpolate between them to determine the final values of $R$. As our predictions for $R$ as a function of $\eta_{\mathrm{R}}$ vary between the simulations performed with different temporal and spatial resolution, we must interpolate between all possible combinations of $R(\eta_{\mathrm{R}}=0.2)$ and $R(\eta_{\mathrm{R}}=0.4)$ to fully realise the final distribution in $R$.

\subsection{The \texorpdfstring{$R_2$}{R2} parameter}
\label{sec: R2 apppendix}

\begin{figure*}
    \centering
    \includegraphics{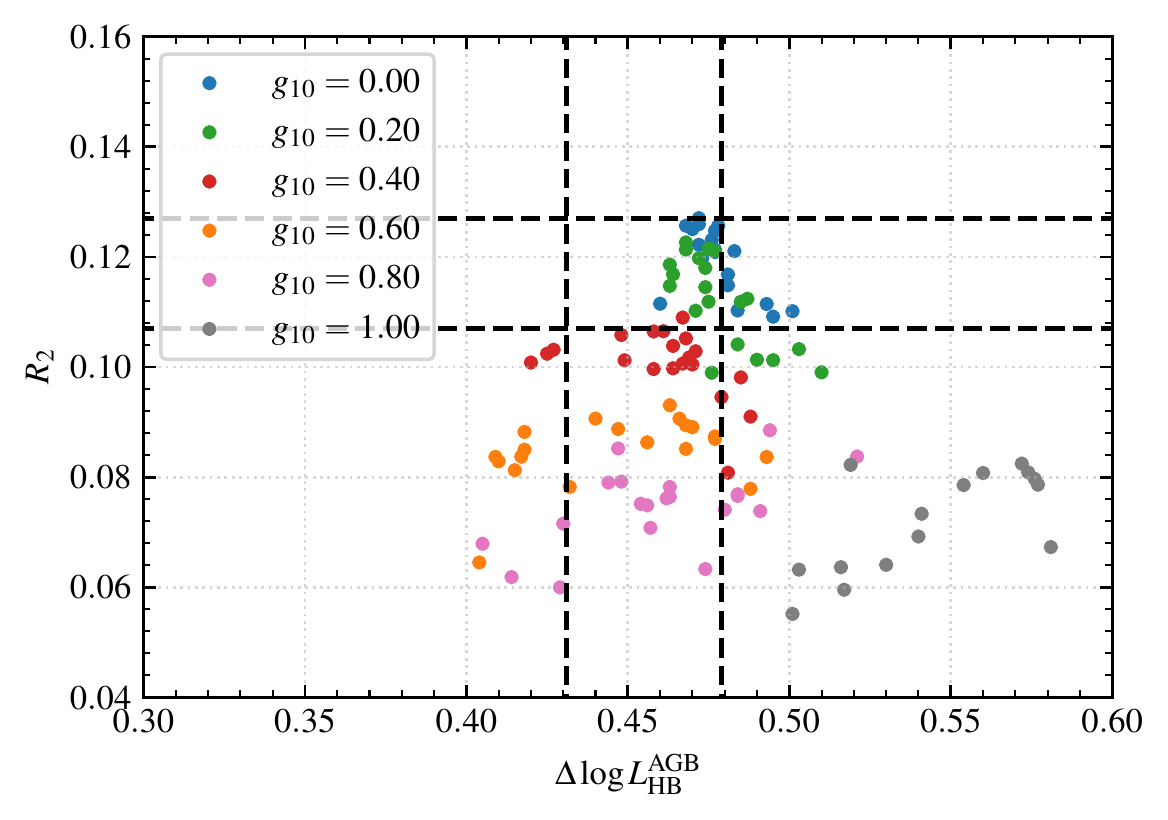}
    \caption{Scatter-plot indicating the values of $R_2$ and $\Delta\log L_{\mathrm{HB}}^{\mathrm{AGB}}$ for each of the simulations performed given standard overshoot with $f_{\mathrm{ov}}=0.001$. Each colour corresponds to a different axion-photon coupling strength, as indicated. The dashed black lines represent the observed 95\% confidence limits for each parameter.}
    \label{fig: R2-DeltaL scatter}
\end{figure*}

In Section \ref{sec: R2parameter} we derive a constraint on $g_{10}$ based on the star cluster count method originally presented in Ref \cite{Lattanzio2}. We chose to frame this discussion around $R_2$ and neglected the relevance of another globular cluster property, the HB-AGB luminosity difference $\Delta\log L_{\mathrm{HB}}^{\mathrm{AGB}}$, which is constrained by observation to be $0.455\pm0.05$ at the $1\sigma$ level \cite{Lattanzio2}.

There are a number of reasons for omitting any discussion of $\Delta\log L_{\mathrm{HB}}^{\mathrm{AGB}}$. Firstly, axion energy-loss primarily affects $R_2$ rather than $\Delta\log L_{\mathrm{HB}}^{\mathrm{AGB}}$. This can be seen Fig. \ref{fig: R2-DeltaL scatter}, which shows the values of $\Delta\log L_{\mathrm{HB}}^{\mathrm{AGB}}$ for the same set of simulations presented in Fig. \ref{fig: R and R2} given $f_{\mathrm{ov}}=0.001$. The colour of each set of dots represents the adopted value of $g_{10}$ and are indicated in the legend.

Although the inclusion of additional energy loss increases the spread of $\Delta\log L_{\mathrm{HB}}^{\mathrm{AGB}}$, the average remains within the observed 95\% C.I. in all cases except $g_{10}=1.0$. This corresponds to values of $R_2$ which are already excluded.

Secondly, calculations of $\Delta\log L_{\mathrm{HB}}^{\mathrm{AGB}}$ are particularly sensitive to variation in input physics. Specifically, Ref \cite{Lattanzio2} estimates the 1 standard deviation uncertainty in calculations in this parameter are $\sigma_{\Delta\log L}^{\mathrm{sys}}=0.04$, considerably larger than the equivalent observed uncertainty of $\sigma_{\Delta\log L}^{\mathrm{obs}}=0.012$, with the most sizeable contributions coming from variation in composition. While we have investigated the effect of $Y_{\mathrm{init}}$ in this work, repeating these simulations for other values of $Z$ was needlessly cumbersome computationally, given the minor effect it has on both $R$ and $R_2$.

The luminosity difference $\Delta\log L_{\mathrm{HB}}^{\mathrm{AGB}}$ can also exhibit strong dependence on the adopted value of
$f_{\mathrm{ov}}$ during the early AGB. In general, the helium burning shell of the E-AGB is radiative and hence convective mixing parameters do not affect its evolution. A notable exception to this occurs when it encounters large composition discontinuities in the intermediate region described in Section \ref{sec: CBs} and Appendix \ref{sec: CHeB CBs}. As the shell passes through these discontinuities, the supply of helium in the region of strongest HeB increases dramatically. This is followed by a brief period of intense HeB activity, before settling back into stable shell burning. These events are known as \textit{gravonuclear loops}, and cause small convective pockets to emerge within the shell.

It was noted in Ref \cite{Lattanzio2} that the presence of convective overshoot \textit{below} such regions causes a decrease in the predicted value of $\Delta\log L_{\mathrm{HB}}^{\mathrm{AGB}}$. While this does not cause major discrepancies for small choices of $f_{\mathrm{ov}}$ (e.g. $0.001$), values of $f_{\mathrm{ov}}=0.01$ or greater typically result in predictions of $\Delta\log L_{\mathrm{HB}}^{\mathrm{AGB}}\approx0.3$. Convective mixing schemes which leave these large discontinuities must therefore wrestle with the additional uncertainty in $\Delta\log L_{\mathrm{HB}}^{\mathrm{AGB}}$ from this source.

These factors demonstrate the difficulty of constraining stellar simulations from $\Delta\log L_{\mathrm{HB}}^{\mathrm{AGB}}$ alone, and motivate why our discussion in Section \ref{sec: R2parameter} focused solely on $R_2$.

\section{Results for Mixing Schemes}
\label{app: Results}
The limits obtained for all convective boundary schemes considered are provided in Tables \ref{tab: SO_SC} (Standard Overshoot and Semi-convection) and \ref{tab: PM_CP} (Convective Premixing and Predictive Mixing).

\begin{table*}
\begin{tabular}{>{\centering}m{0.1\textwidth}>{\centering}m{0.1\textwidth}>{\centering\arraybackslash}m{0.1\textwidth}} 
\multicolumn{3}{c}{Standard Overshoot} \\
\hline\hline
$f_{\mathrm{ov}}$ & $g_{10}^R$ & $g_{10}^{R_2}$  \\ 
\hline
0.0001            & 0.63     & 0.39            \\
0.001             & 0.65     & 0.43           \\
0.01              & 0.72     & 0.28           \\
\hline\hline
\end{tabular}
\hspace{2.5cm}
\begin{tabular}{>{\centering}m{0.08\textwidth}>{\centering}m{0.08\textwidth}>{\centering}m{0.08\textwidth}>{\centering\arraybackslash}m{0.08\textwidth}} 
\multicolumn{4}{c}{Semiconvection} \\
\hline\hline
$f_{\mathrm{ov}}$      & $\alpha_{\mathrm{sc}}$ & $g_{10}^R$ & $g_{10}^{R_2}$  \\ 
\hline
\multirow{3}{*}{0.001} & 1.0                    & 0.53      & 0.47           \\
                       & 0.1                    & 0.51      & 0.43           \\
                       & 0.001                  & 0.56      & 0.28           \\
\hline
\multirow{2}{*}{0.01}  & 1.0                    & 0.60      & 0.37           \\
                       & 0.1                    & 0.63      & 0.37           \\
\hline\hline
\end{tabular}
\caption{\label{tab: SO_SC} These tables show the values of the constraint we would obtain using different values of the parameters for Standard Overshoot (left table) and Semiconvection (right table).}
\end{table*}

\begin{table*}
\centering
\begin{tabular}{>{\centering}m{0.1\textwidth}>{\centering}m{0.1\textwidth}>{\centering\arraybackslash}m{0.1\textwidth}} 
\multicolumn{3}{c}{Convective Premixing} \\
\hline\hline
Resist He & \multirow{2}{*}{$g_{10}^R$} & \multirow{2}{*}{$g_{10}^{R_2}$}  \\
increase & & \\
\hline
Yes                & 0.69      & 0.37           \\
No                 & 0.77      & 0.23           \\
\hline\hline
\end{tabular}
\hspace{2.5cm}
\begin{tabular}{>{\centering}m{0.08\textwidth}>{\centering}m{0.08\textwidth}>{\centering}m{0.08\textwidth}>{\centering\arraybackslash}m{0.08\textwidth}} 
\multicolumn{4}{c}{Predictive Mixing} \\
\hline\hline
$\delta_{\mathrm{PM}}$ & $\alpha_{\mathrm{i}}$ & $g_{10}^R$ & $g_{10}^{R_2}$  \\ 
\hline
\multirow{4}{*}{0.005} & 0.21                  & 0.52      & 0.11           \\
                       & 0.42                  & 0.55      & 0.17           \\
                       & 0.83                  & 0.58      & 0.19           \\
                       & No    & \multirow{2}{*}{0.77}      & \multirow{2}{*}{0.28}           \\
                       & restriction & & \\
\hline
\multirow{2}{*}{0.01}  & 0.42                  & 0.53      & 0.21           \\
                       & No        & \multirow{2}{*}{0.80}      & \multirow{2}{*}{0.29}           \\
                       & restriction & & \\
\hline\hline
\end{tabular}
\caption{\label{tab: PM_CP} These tables show the values of the constraint we would obtain using different values of the parameters for Convective Premixing (left table) and Predictive Mixing (right table).}
\end{table*}

\acknowledgments
This work was supported in part by the Australian Research Council through the ARC Centre of Excellence for Dark Matter Particle Physics, CE200100008 and the Australian Government Research Training Program Scholarship initiative. We would like to sincerely thank Amanda Karakas, Simon Campbell and John Lattanzio for their advice regarding the $R_2$-parameter constraint, as well as Oscar Straniero and Alessandro Mirizzi for helpful comments.

\bibliographystyle{JHEP}
\bibliography{bibliography}
\end{document}